\documentclass[12pt]{article}

\usepackage{newtxtext,newtxmath}

\usepackage{graphicx}

\usepackage[letterpaper,margin=1in]{geometry}

\renewenvironment{abstract}
	{\quotation}
	{\endquotation}

\date{}

\makeatletter
\renewcommand{\fnum@figure}{\textbf{Figure \thefigure}}
\renewcommand{\fnum@table}{\textbf{Table \thetable}}

\makeatother

\usepackage{scicite}

\usepackage{url}

\def\scititle{
	Evidence for a polar circumbinary exoplanet orbiting a pair of
eclipsing brown dwarfs
}

\title{{\bfseries \boldmath \scititle}}

\author{
	Thomas~A.~Baycroft$^{1\ast}$,
	Lalitha~Sairam$^{1,2}$,
	Amaury~H.M.J.~Triaud$^{1}$\and
        Alexandre~C.M.~Correia$^{3,4}$\and
	\small$^{1}$School of Physics and Astronomy, University of Birmingham,
Edgbaston, Birmingham, B15 2TT, United Kingdom.\and
	\small$^{2}$Institute of Astronomy, University of Cambridge, Madingley road,
Cambridge, CB3 0HA, United Kingdom.\and
        \small$^{3}$CFisUC, Departamento de Fisica, Universidade de Coimbra, 3004-516
Coimbra, Portugal.\and
        \small$^{4}$IMCCE, UMR8028 CNRS, Observatoire de Paris, PSL Universite, 77
avenue Denfert-Rochereau, 75014 , Paris, France.\and
	\small$^\ast$Corresponding author. Email: txb187@bham.ac.uk\and
}

\begin{document} 

\maketitle

\begin{abstract} \bfseries \boldmath
One notable example of exoplanet diversity is the population of \textit{circumbinary planets}, which orbit around both stars of a binary star system. There are so far only 16 known circumbinary exoplanets, all of which lie in the same orbital plane as the host binary. Suggestions exist that circumbinary planets could also exist on orbits highly inclined to the binary, close to \(90^{\circ}\), \textit{polar} orbits. No such planets have been found yet but polar circumbinary gas and debris discs have been observed and if these were to form planets then those would be left on a polar orbit. We report strong evidence for a polar circumbinary exoplanet, which orbits a close pair of brown dwarfs which are on an eccentric orbit. We use radial-velocities to measure a retrograde apsidal precession for the binary, and show that this can only be attributed to the presence of a polar planet.
\end{abstract}

\section{Introduction}

Sixteen circumbinary exoplanets have been detected to-date. The majority were identified using the transit method with \textit{Kepler} \cite{doyle_kepler-16_2011} and \textit{TESS} \cite{kostov_toi-1338_2020}. Given the sample of 12 transiting circumbinary planets found by \textit{Kepler}, it is argued that the main population of circumbinary planets must be close to coplanar with the binary (with mutual inclinations $\Delta i < 5^\circ$) \cite{armstrong_abundance_2014}, otherwise the occurrence rate of circumbinary planets would be greater than that of similar-sized planets orbiting single stars. (Preliminary results from the BEBOP (Binaries Escorted By Orbiting Planets) search for circumbinary planets using radial velocities are broadly consistent with those results \cite{baycroft_progress_2024}).\\

Despite this observational evidence, there may still exist a comparatively small and yet-undetected population of circumbinary planets occupying misaligned \cite{martin_planets_2014,childs_misalignment_2022} or polar \cite{childs_formation_2021} orbits. Polar circumbinary orbits are stable, even very close to the binary, if the binary is eccentric enough and the third body's orbital plane lies perpendicular to the inner binary's line of apsides \cite{doolin_dynamics_2011,chen_orbital_2019,chen_polar_2020}. A circumbinary orbit can exist in one of two possible states. With sufficient inclination to an eccentric binary the orbit will {\it librate} about the binary's eccentricity vector. The critical inclination above which this happens depends on the binary eccentricity \cite{doolin_dynamics_2011}, if the inclination is below this limit then the orbit will instead {\it circulate} about the binary's angular momentum vector. \\

The existence of circumbinary planets on polar orbits, while exotic and seemingly unlikely, has a theoretical and observational basis, and has been postulated before \cite{schneider_occultations_1994, martin_planets_2014}. Circumbinary protoplanetary and debris discs in a polar orientation have been observed in a few systems \cite{kennedy_99_2012,kennedy_circumbinary_2019}, and various mechanisms exist to form misaligned discs \cite{bonnell_fragmentation_1992,offner_formation_2010,bate_diversity_2018,nealon_flyby-induced_2020} as well as to form planets from these discs that would therefore be misaligned or polar \cite{childs_formation_2021,childs_misalignment_2022}. One of the polar discs also displays evidence for dust growth, indicating planet formation within polar discs is feasible \cite{kennedy_circumbinary_2019}. While no polar circumbinary planets have been found so far, it has been suggested that AC Her a post-asymptotic giant branch binary star, which has a polar circumbinary disc \cite{martin_ac_2023}, may also host a third body. The disc in that case is truncated and could be explained by a companion orbiting the binary interior to the disc \cite{kluska_population_2022} in an assumed polar orbit. Polar orbits are also fairly frequently encountered for hot Jupiters orbiting single stars \cite{triaud_rossiter-mclaughlin_2018,albrecht_preponderance_2021}. It has been suggested that polar circumbinary planets where the inner binary has subsequently merged could be a possible origin of polar planets around single stars \cite{chen_origin_2024}.\\

The radial velocity method has now detected three circumbinary planets: detecting Kepler-16b \cite{triaud_bebop_2022}, confirming the detection of and improving the physical and orbital parameters of TIC\,172900988\,b \cite{sairam_new_2024}, and the first stand-alone radial velocity discovery of BEBOP-1c \cite{standing_radial-velocity_2023} (the inner planet TOI-1338b having been found in transit \cite{kostov_toi-1338_2020}). These are all detections made by measuring the reflex orbital motion of the centre-of-mass of the system caused by a planet. A third body such as a circumbinary planet also has a dynamical influence on the orbit of the binary, most notably on its apsidal precession rate. Measurements of the apsidal precession rate have been used to constrain the masses of some of the \textit{Kepler} transiting circumbinary planets \cite{kostov_kepler-1647b_2016} and can in principle be used on their own to infer the presence of a planet within radial-velocity data.\\

2MASS J15104786-2818174 (hearafter 2M1510) is a double-lined eclipsing binary composed of two equal-mass brown dwarfs \cite{triaud_eclipsing_2020} of mass 0.0331 and 0.0332 \(M_{\rm \odot}\) respectively. The orbital geometry of the binary results in a single eclipse. This is due to the combination of a high eccentricity, a very slight inclination with respect to the line-of-sight, and the line-of-apsides being along the line-of-sight. The system has been found to be a kinematic member of the \(45\pm5\) Myr old Argus moving group, making it the second of only two known young eclipsing double-lined brown dwarf binaries, important objects for calibrating brown dwarf evolutionary models \cite{burrows_theory_2001,stassun_discovery_2006,triaud_eclipsing_2020}. The binary has an orbital period of 20.9 days, an eccentricity of 0.36, and there is also a visual brown dwarf tertiary companion, at a projected separation of \(\sim250\) AU from the central, eclipsing pair \cite{triaud_eclipsing_2020}.\\

Archival and newly obtained radial velocities of 2M1510 exist, observed with the UVES (Ultraviolet and Visual Echelle Spectrograph) instrument on the European Southern Observatory (ESO)'s  Very Large Telescope (VLT). In this work we analyse the radial velocities and detect strong evidence for a retrograde apsidal precession. We interpret this as a sign of a polar orbiting circumbinary planet, one of the few possible causes for a retrograde precession \cite{zhang_distinguishing_2019}.

\section{Results}

\subsection*{Retrograde apsidal precession}

\begin{figure}[h]
\centering
\includegraphics[width=0.96\textwidth]{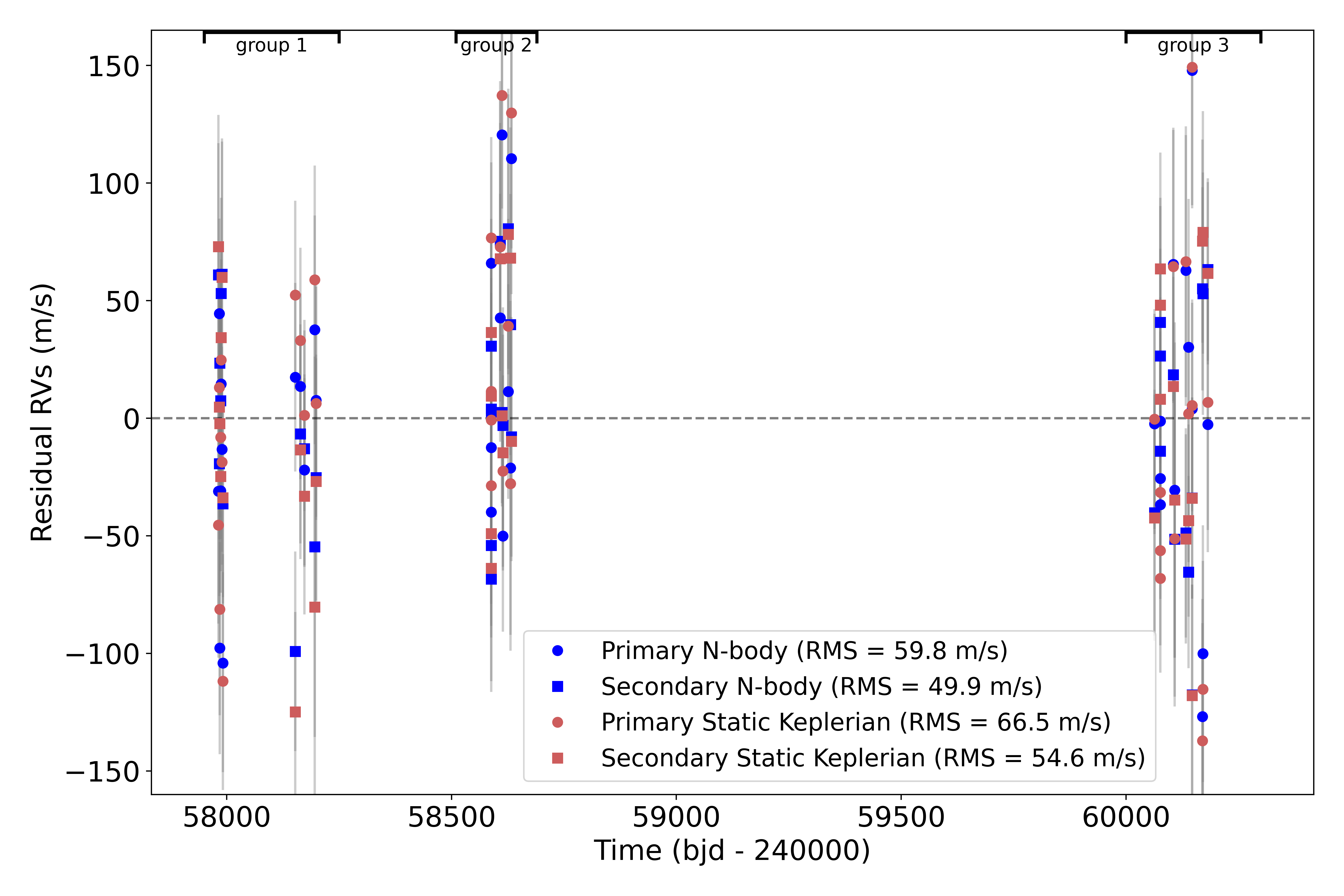}
\caption{\textbf{Comparison of N-body residuals and static-Keplerian residuals.} The division into different groups used to see the evolution of \(\dot{\omega}\) is shown. The RMS of the residuals for each dataset is also listed.}
\label{fig:resids}
\end{figure}

We perform a first fit to the radial velocities using {\tt kima} \cite{faria_kima_2018}, an exoplanetary tool that utilises nested sampling to search for an arbitrary number of Keplerian signals, and explore their parameter space. We use {\tt kima} to fit the radial velocities of both brown dwarfs simultaneously assuming a Keplerian model, but also adding an apsidal precession rate \(\dot{\omega}\) \cite{baycroft_improving_2023}, the most important Newtonian perturbation. This analysis leads to improved constraints on the parameters of the binary, which we present in the first column of Table \ref{tab:params}, and importantly, to a measure of \(\dot{\omega} = -343\pm126~{\rm "/yr}\), corresponding to a negative apsidal precession rate, which is detected with \(99.7 \%\) confidence. The radial velocity residuals and the posterior distribution on the precession rate are shown in Fig~\ref{fig:resids}~and Fig~\ref{fig:prisec}. No Doppler reflex signal consistent with a circumbinary planet signal is detected in the radial velocities.\\

A negative (i.e. retrograde) apsidal precision is a rare and immediately noticeable result. Typical effects inducing an apsidal precession (General relativity, tidal distortion) would {\it always} result in a prograde (i.e. positive) apsidal precession rate \cite{baycroft_improving_2023}. For 2M1510, these effects combined result in a precession \({\dot \omega}_{\rm bin} \lesssim 4{\rm ~"~yr^{-1}}\) much smaller in magnitude than what is measured. Apsidal precession induced by a third body is also only prograde if the companion is coplanar with the binary. However, a companion on a highly misaligned/polar orbit could cause a retrograde apsidal motion \cite{zhang_distinguishing_2019}. The most stable circumbinary polar orbit is perpendicular to the eclipsing binary's apsides \cite{doolin_dynamics_2011,chen_orbital_2019}, and because of the binary's orbital parameters ($i_{\rm bin} \sim 90^\circ, \omega_{\rm bin} \sim 270^\circ$) mean that its apsides are along the line-of-sight, the polar planet would therefore appear face-on from our point of view, which is consistent with the absence of Doppler reflex motion observed in the radial-velocity timeseries. Alternative explanations are explored in section~\ref{sec:discussion} but none can explain the measurement or retrograde precession. In the absence of any other viable mechanism to cause it, the precession must therefore be due to a perturbing third body on a polar orbit.\\

The companion responsible for the retrograde apsidal precession is most likely planetary. There are only two sets of absorption lines in the UVES spectra and the extracted radial velocities based on a 2-spectrum model have a small scatter (\(<60\) m/s), so the third body must be much less luminous than the inner brown dwarf pair. Since their individual masses are around \(35 M_{\rm Jup}\), a fainter, inclined companion is very likely be planetary in nature.

\subsection*{N-body analysis}\label{sec:nbody}

To assess what planet configurations could cause this precession rate we fit the radial velocity data using an N-body fit (more details in section \ref{sec:methods}. The binary parameters from this fit are shown in Table \ref{tab:params}.
\\

\begin{figure}[h]
\centering
\includegraphics[width=0.9\textwidth]{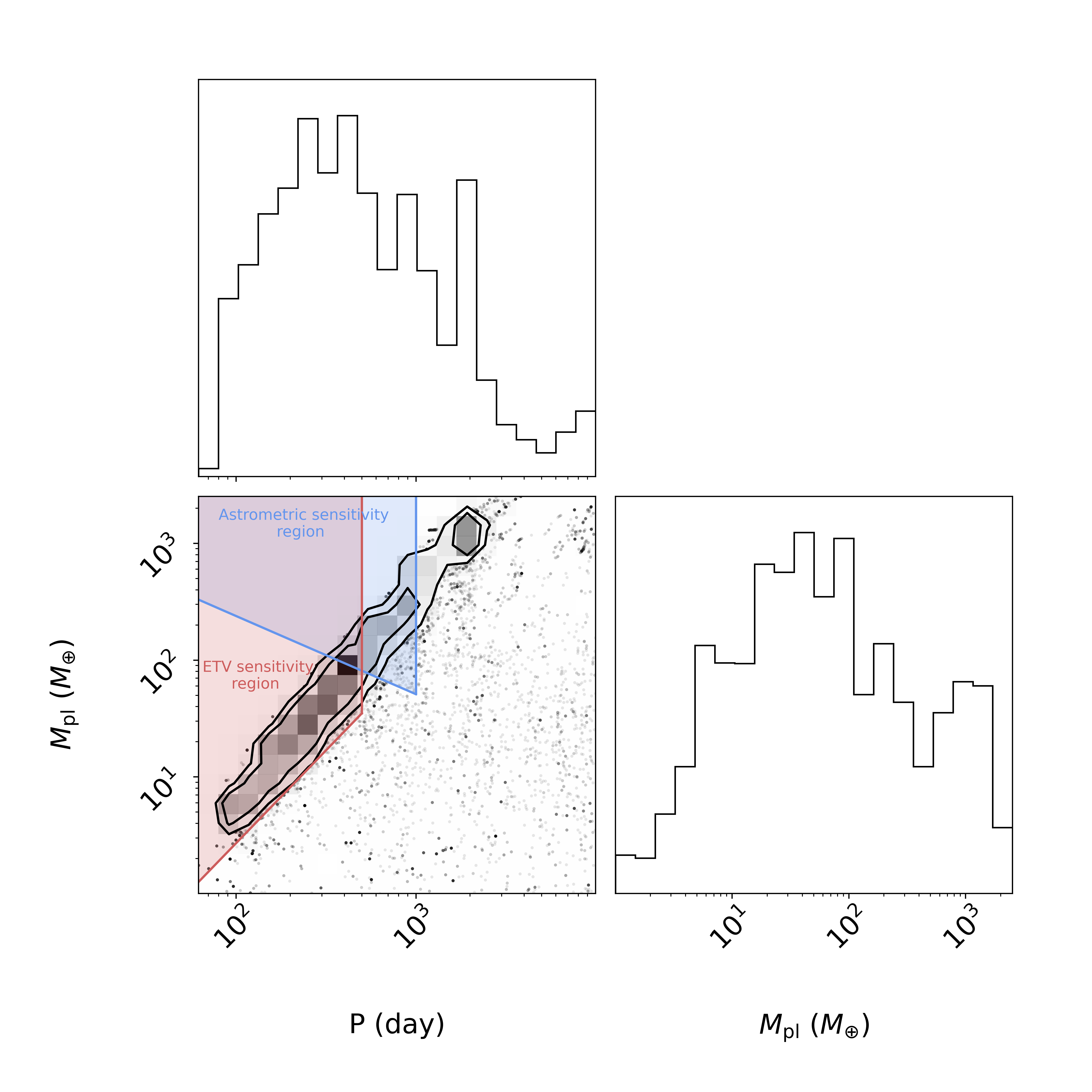}
\caption{\textbf{Planet masses and periods} Corner plot of Periods and masses consistent with the induced precession rate from an N-body fit. Regions of sensitivity are shown for: 1) a three year astrometric campaign with the VLT in blue and 2) two years of eclipse timing monitoring with 1 minute timing precision (as was achieved with SPECULOOS) in red.}\label{fig:corner}
\end{figure}

Fig~\ref{fig:corner} shows the constraints on the mass and orbital period of the third body from the N-body fit. Since the radial velocities are not directly sensitive to that third object, neither the mass nor the orbital period is individually constrained (testing with different priors shows that we remain prior-dominated). However, the relationship between mass and period is constrained through the the apsidal precession rate. This leads to a ``wedge" of masses and periods where the N-body integrator reproduces the observed radial-velocities. Most circumbinary planets are detected right outside the unstable region surrounding the inner binary \cite{doolin_dynamics_2011, martin_planets_2014}. Should this be the case for 2M1510\,b, the polar planet, then its parameters would be of order $P \approx 100$ days and $M \approx 10 \,M_{\rm \oplus}$. If instead $P \approx 400$ days, then $M \approx 100\,M_{\rm \oplus}$. \\

\begin{figure}[h]
\includegraphics[width=0.98\textwidth]{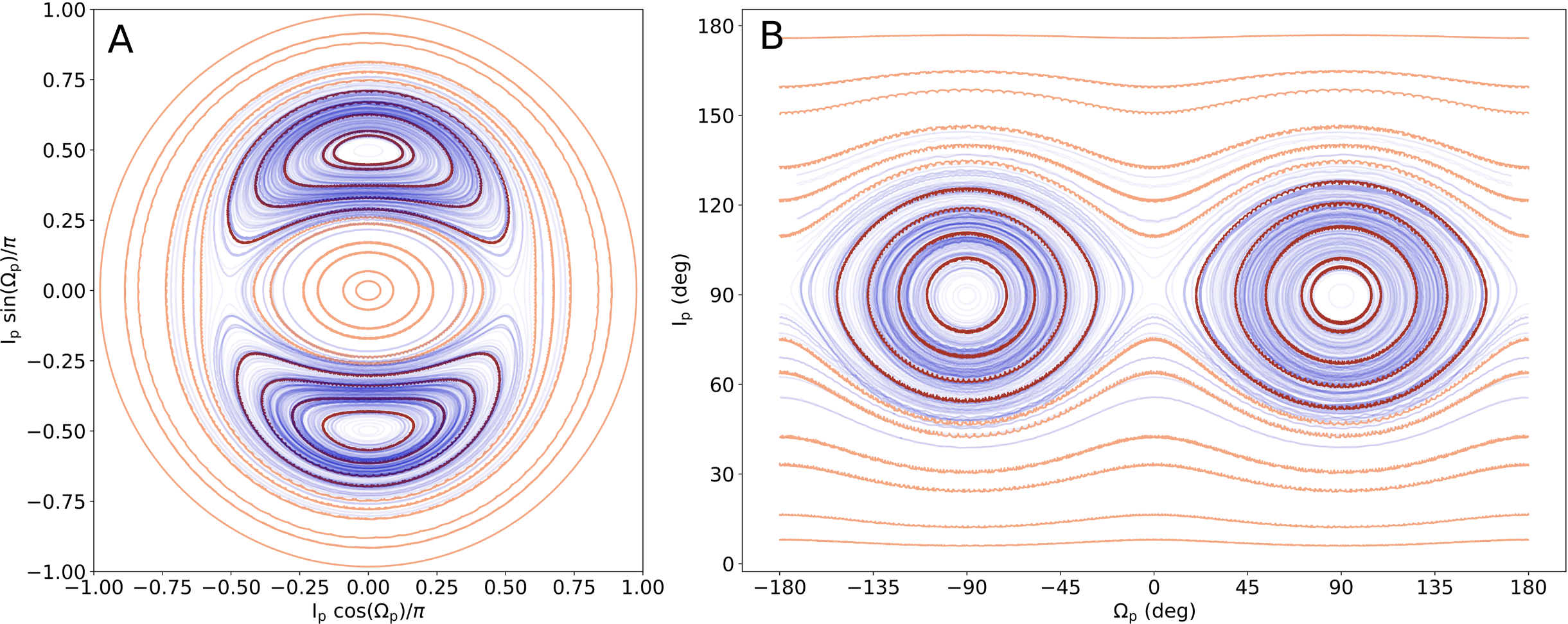}
\caption{\textbf{Level curves of the Hamiltonian for 2M1510 with the colours
denoting the regions of libration and circulation.} These are trajectories of test particles orbiting the binary. The red trajectories show the regions of libration and the orange trajectories the regions of circulation. The 584 stable orbits of 600 randomly drawn simulations from the posterior of the N-body fit are integrated and the trajectories overplotted in blue. Note that since the simulations shown in blue are drawn from the N-body fit they are for massive objects not test particles. Two different projections are shown.}\label{fig:IWpost}
\end{figure}

The planet is indeed a polar planet in the librating regime. Fig~\ref{fig:IWpost} shows the energy levels of the Hamiltonian with the posteriors from the N-body fit plotted on top. We show the islands of libration and circulation of a test particle within the 2M1510 binary, with red showing circulation and orange depicting libration. We take a random sample of 600 posteriors from the run restricted to periods \(<300\) days. 16 of these are unstable (the orbital period of the planet changes by a factor of 1.5 within 400,000 days) and of the remaining 584 posterior samples, 96\% are in the librating regime.
\\

\section{Discussion}\label{sec:discussion}

We discuss alternative explanations for the retrograde apsidal precession measurement, ultimately finding no suitable explanation other than a polar planet. We then discuss methods to follow-up and confirm the planet and highlight a point on the nomenclature.

\subsection*{Alternative explanations}\label{sec:other}

In this section we explore alternative explanations for the measured \(\dot{\omega}\). Each of the following subsections is given one alternative hypothesis as a title, this hypothesis is then discussed. Overall we find that there is no suitable explanation other than a polar-orbiting planet.

\subsubsection*{The precession is a false-positive} The evidence for this candidate planet is predicated on the measurement of a retrograde apsidal precession. This measurement is over the detection thresholds and we now perform some checks of the reliability by re-analysing with {\tt kima}, using different portions of the data.\\

\begin{figure}[h]
\centering
\includegraphics[width=0.98\textwidth]{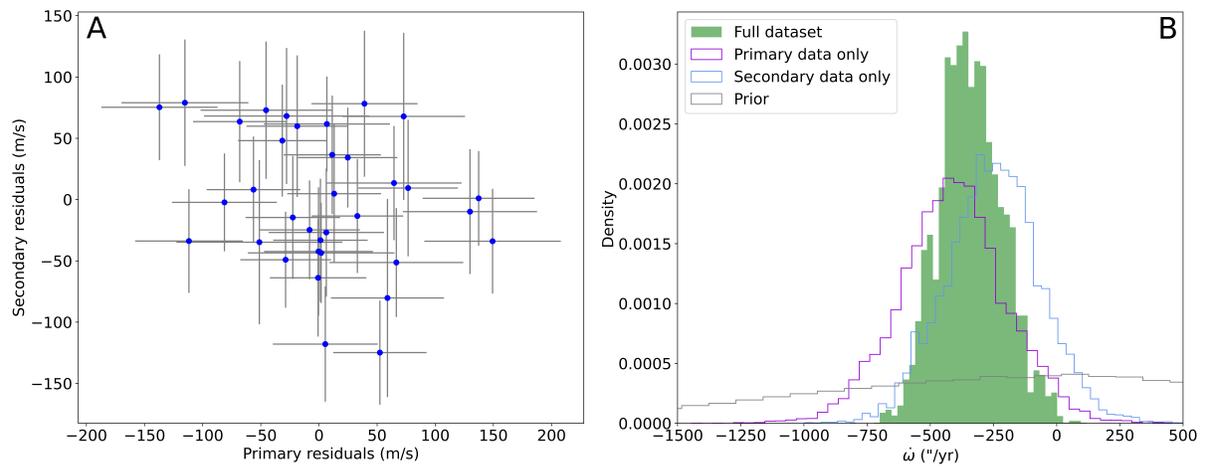}
\caption{\textbf{Independent analysis of primary and secondary radial velocity data.} Panel A (left) shows the lack of correlation between the residuals of the primary and secondary. Panel B (right) shows the posterior distribution of \(\dot{\omega}\) for the analyses of the primary, the secondary, and the full dataset compared to the prior distribution.}
\label{fig:prisec}
\end{figure}

First, we separate the data into the radial velocities from each of the primary and secondary brown dwarfs, and analyse each separately. The analysis produces independent measurements of \(\dot \omega_{\rm bin} =-410\pm200\) "\,yr$^{-1}$ for the primary and  \(\dot \omega_{\rm bin} =-250\pm180\) "\,yr$^{-1}$ for the secondary. The posterior distributions for these are shown in the right hand panel of Fig \ref{fig:prisec}. Both results are consistent with one another. The left hand panel of Fig \ref{fig:prisec} shows the relationship between the residuals of the primary and the residuals of the secondary when fit with a static (non-precessing) Keplerian. The lack of correlation shows that the two datasets are indeed independent measures of the apsidal precession rate, so the detection is not due to a spurious structure in the residuals of one of the two components.\\

Second, we separate the data in time. The data was taken in 3 different groups (see Fig \ref{fig:resids}). We analyse the first two groups (covering a baseline of 650 days) and the second two groups (covering a baseline of 1600 days). The apsidal precession measurements are \(+860\pm650\) and \(-460 \pm 350\) "\,yr$^{-1}$ respectively. While the first measurement is prograde, it is \(<2\sigma\) from the retrograde value based on analysing the full dataset, and it is taken over a very small time baseline from which conclusions about the long-term precession rate are hard to extract. This shows that the measurement is most strongly reliant on the final group of data, but this is expected as this group extends the baseline the most and a long baseline is vital to a good measurement of the apsidal precession rate of a binary. The fact that the measurement refines in precision and becomes more negative when more data is included gives us confidence the signal is real. However, further radial velocity measurements extending the baseline would be very helpful in constraining the negative $\dot \omega$, refining its precision. This is also important to better constrain the possible mass and period combinations for the polar planet, the current constraint being shown in Fig~\ref{fig:corner}.

\subsubsection*{The companion is non-planetary} The companion causing this precession must be planetary mass. Unfortunately, we cannot put an upper limit on the companion's mass based on an N-body fit as there will always be perfectly face-on orbits of very massive bodies on long-period orbits that can induce the apsidal precession but create no radial velocity signature. However, it is unlikely that the third body is itself a brown dwarf, as it would have already been noticed. Prior to 2M1510 being known to be eclipsing, \cite{gizis_brown_2002} had speculated decades ahead this system was an equal brightness binary, based on its overall luminosity. To avoid photometric and spectroscopic detection, any companion to the binary would have to be much weaker. Since both components of the binary are themselves low-mass brown dwarfs ($\sim 30\,\rm M_{jup})$, this only leaves the planetary realm as a reasonable solution. 

\subsubsection*{The distant tertiary is producing the precession}
2M1510 AB, the eclipsing binary brown dwarf in question here, has a visual companion of the same magnitude (2M1510 C). It was shown that with the projected separation of 250 AU and an orbital period of \(\sim 11,000\) years, the tertiary is unable to have caused the binary to reach its current state through high-eccentricity migration \cite{triaud_eclipsing_2020}. We use equation A5. from \cite{baycroft_improving_2023} to calculate the precession rate due to the tertiary and find that the effect is six orders of magnitude too small. The distant tertiary cannot be the cause of the apsidal precession.

\subsubsection*{Proper-motion changing the viewing angle} An apparent apsidal precession can occur for a high proper-motion system due to the viewing angle changing \cite{rafikov_stellar_2009}. This is only important for very high proper-motion systems, for 2M1510 we calculate an upper limit of \(|\dot{\omega}| < 0.13 {\rm\,"/yr}\), so this effect is negligible here.

\subsubsection*{A spin--orbit misalignment} Retrograde precession can also be induced if the binary has a spin--orbit misalignment, as is the case in DI Her \cite{albrecht_misaligned_2009} where the retrograde precession induced in this system made its measured precession, while still prograde, much smaller than is predicted by GR alone. However, this effect is negligible here too.

We utilise eq. (\ref{eq:wdot}) derived below. The eccentricity (\(e\)), semi-major-axis (\(a\)), and the masses are observables that we have measured or derivable from them. \cite{triaud_eclipsing_2020} measured the inclination (\(I\)), and constrained the rotational periods of the stars to be between 20-30 hours. We assume 20 hours rotation period for both brown dwarfs to be conservative. We use Love numbers (\(k_{2,i}\)) of 0.4, corresponding to apsidal constants of 0.2 slightly more conservative that the value of 0.143 used for a similar brown dwarf system \cite{heller_tidal_2010}. We draw \(1,000,000\) randomisations of the binary longitude of ascending node (\(\Omega\)), and the angles \(\varepsilon_i\) and \(\varphi_i\). From these we calculate that \(\dot{\omega}>-2.15\) "/yr so the spin-induced apsidal precession is two orders of magnitude too small to cause the measured effect. 

\subsubsection*{A polar disc} Apsidal precession induced by a third body is a secular effect. This could equally be produced by a highly inclined circumbinary disc. However, while the binary is relatively young at $\sim 45 \,\rm Myr$ \cite{triaud_eclipsing_2020} it is old enough for the protoplanetary disc to have dissipated \cite{mamajek_initial_2009}, and no infrared excess is noticed \cite{triaud_eclipsing_2020}. Any remaining and undetected disc would presumably be a debris disc. Distinguishing a massive close-in debris disc from a single planet, or from multiple planets is not possible. All contribute to apsidal precession, and we chose to favour the simplest version of a single perturbing body.

\subsection*{Follow-up}

Next steps should extend the baseline that is covered with more radial velocities to improve the precision on the apsidal precession rate. We estimate that if 20 radial velocity measurements are taken in the next observing period (ESO P115) with UVES this would lead to a precision on \(\dot{\omega}\) of \(\sim90\) "\,yr$^{-1}$ and another 20 measurements the next observing period (ESO P117) to a precision of \(\sim75\) "\,yr$^{-1}$. However even with more data, we expect the mass-period degeneracy will remain very difficult to break. A few additional ways of confirming the planet and its polar nature and constraining the planet parameters are possible. \\

Firstly, measuring the binary's eclipse times (and depths) precisely could in principle detect the planet's perturbation on the binary and constrain their mutual inclination as in \cite{goldberg_5mjup_2023}. The eclipse obtained with one of the SPECULOOS (Search for Planetary transits EClipsing ULtra cOOl Stars) telescopes in \cite{triaud_eclipsing_2020} has a precision just under 1 minute. Using random draws from the posterior sample of the N-body fit, we calculate the eclipse times over 2 years and measure the amplitude of the O-C residuals. 65\% of the draws had an amplitude \(>5\) mins, and 88\% an amplitude \(>1\) min. These are the dynamical Eclipse Timing Variations (ETVs). We also test whether the Light Travel Time Effect (LTTE) ETVs would be detectable using equation 26 from \cite{borkovits_transit_2011}, however even in the best case with the planet as inclined to the line-of-sight as it can possibly be, this would produce an ETV amplitude of \(\sim 20\) s. The dynamical ETVs are a particularly promising way to confirm this planet, and two years of eclipses at 1 minute precision would be sufficient to confirm or rule-out most of the parameter space consistent with the planet. This does however, assume that most of the eclipses over the two years are actually observed and does not take into account difficulties with observations due to daytime or bad weather.\\

Secondly, a detailed astrometric study, such as those done using FORS2 on the VLT by \cite{sahlmann_astrometric_2014} could reveal the signature of such a planet, which is expected on a face-on orbit. \cite{lazorenko_precision_2009} find a typical sensitivity of around 50 \(\mu\)as, this would be equivalent to a Saturn mass planet on a 400 day orbit around our binary brown dwarf, and is therefore a promising follow-up technique which will be sensitive to gas giants on orbits of intermediate separation. At a Gmag of 17.5, Gaia astrometry will not be sensitive enough to constrain a planet as well as ground-based astrometry. \\

A third option is the direct-imaging method where a planet can be detected from its formation's residual heat, using an adaptive optics system to distinguish it from the glare of its host. At its best, the method can detect an object with a flux ratio $\Delta$ F$ \sim \,10^{-5}$ at an angular separation of $0.2"$ \cite{chomez_preparation_2023}. At the distance of 2M1510, this corresponds to a projected orbital separations $>7.3$AU (which is $\sim 27\,000\, \rm days$). At this separation a companion would have to have a mass of \(\sim 0.7 {\,\rm M_{\odot}}\) to induce the required apsidal precession. Direct imaging will therefore not be sensitive to the planet causing the precession, but it could be sensitive to other planets on more distant orbits. Assuming a system age of 50 Myr \cite{triaud_eclipsing_2020}, and using the Baraffe models (COND03)\cite{baraffe_evolutionary_2003,baraffe_new_2015} we estimate a $3.4\rm \,M_{Jup}$ planet could be detectable in this system if at $7.3\,\rm AU$. At further orbital distances, sensitivity improves to $1.7\rm \,M_{Jup}$.

The regions of sensitivity of a 3 year campaign of precise ground-based astrometry, and of a 2 year complete set of eclipse measurements are highlighted in Fig~\ref{fig:corner}.

\subsection*{Nomenclature}

\begin{figure}[h]
\centering
\includegraphics[width=0.96\textwidth]{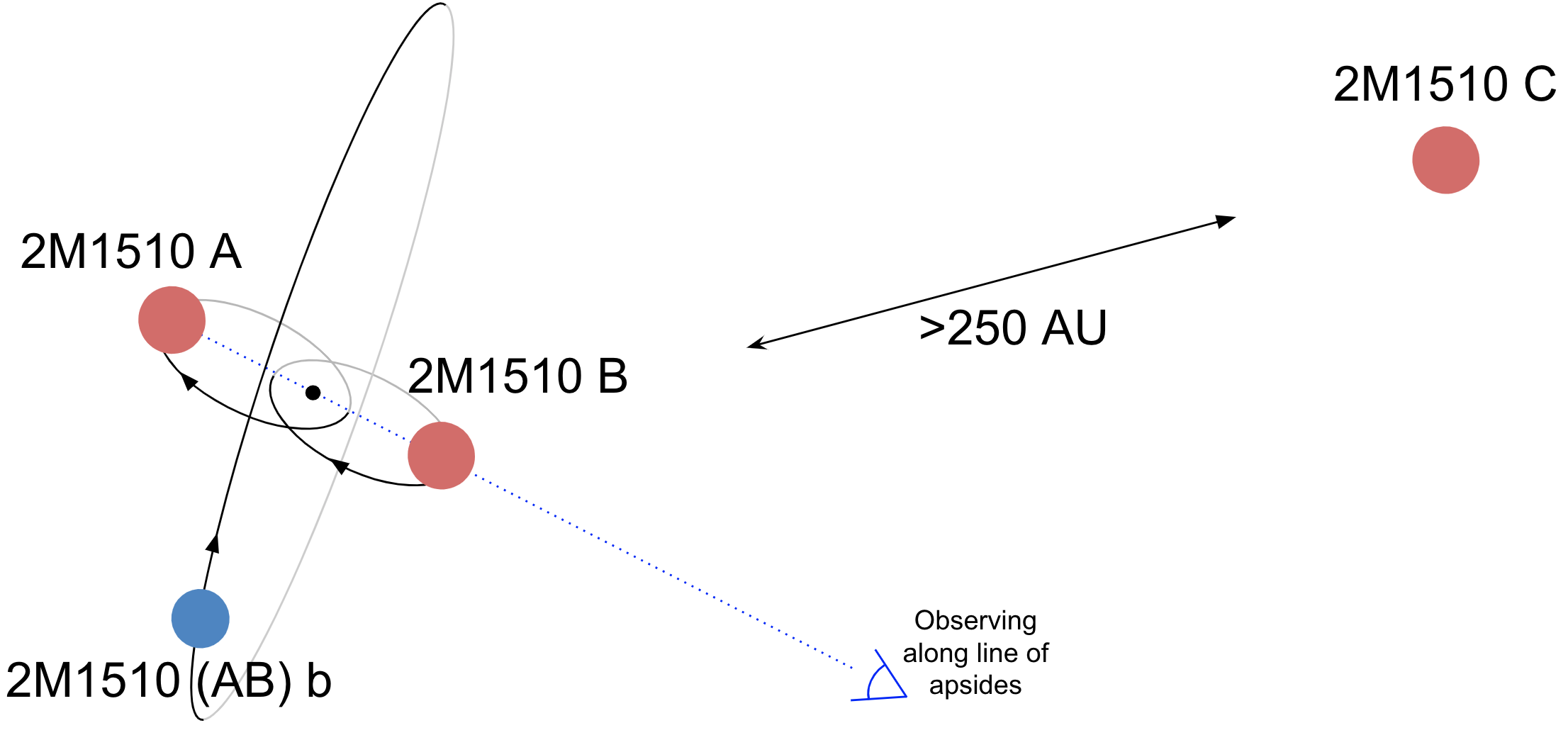}
\caption{\textbf{Configuration of 2M1510 and naming convention for the various bodies.} Brown dwarfs are in red and the planet is in blue. Direction to earth relative to the binary is shown.}\label{fig:config}
\end{figure}

We clarify the naming convention that we are using. While the discovery paper for the eclipsing binary \cite{triaud_eclipsing_2020} referred to the eclipsing binary as 2M1510 A and the distant companion as 2M1510 B. Instead we refer the the two components of the eclipsing binary as 2M1510 A and 2M1510 B; the visual companion is 2M1510 C; the eclipsing binary as a pair of stars is 2M1510 AB; and the planet is 2M1510 (AB)b though we simplify and refer to it as 2M1510 b. Fig~\ref{fig:config} shows the configuration of the system and shows the naming convention we have just described.

\section{Materials and Methods}\label{sec:methods}

\subsection*{Observations and radial velocities}

35 spectra were obtained with the UVES instrument \cite{dekker_design_2000} on the VLT between 2017-08-16 and 2023-08-25 (Prog.ID 299.C-5046, 2100.C-5024 and 0103.C-0042, PI Triaud). All observations used the UVES instrument on UT2-Kueyen. Of these, 13 are recent UVES observations obtained under Prog.ID 111.24ZA.001 (PI Sairam), greatly extending the timespan of observation. 
The spectra were reduced with the standard ESO pipelines for UVES.
\\

We extracted the radial velocities from the spectra using the {\tt DOLBY}-SD method (previously known as SD-GP) presented in \cite{sairam_new_2024}, and inspired by \cite{czekala_disentangling_2017}. {\tt DOLBY}-SD  uses Gaussian Processes to disentangle both spectral components from one another, and calculate precise and accurate radial velocity measurements. This method treats the intrinsic stellar spectra as realisation of Gaussian process with a Mat\'ern kernel. By modelling the radial velocity shift of each star in the binary system due to their orbital motion, the Gaussian process allows us to deconvolve the composite spectrum of the binary system into the individual spectra of each star. We divided the observed spectrum into smaller wavelength chunks to make the Gaussian process calculations tractable. We used the independent sets of hyperparameters to model the spectra of each star within each chunk, accounting for potential differences in the spectral characteristics of each star. We employed Markov Chain Monte-Carlo (MCMC) methods to explore the posterior distribution of both the radial velocities and the Gaussian process hyperparameters. Finally, we combined the radial velocities from each chunk using a weighted average, with weights determined by the uncertainties from each chunk.  
\\

\begin{figure}[h]
\centering
\includegraphics[width=0.96\textwidth]{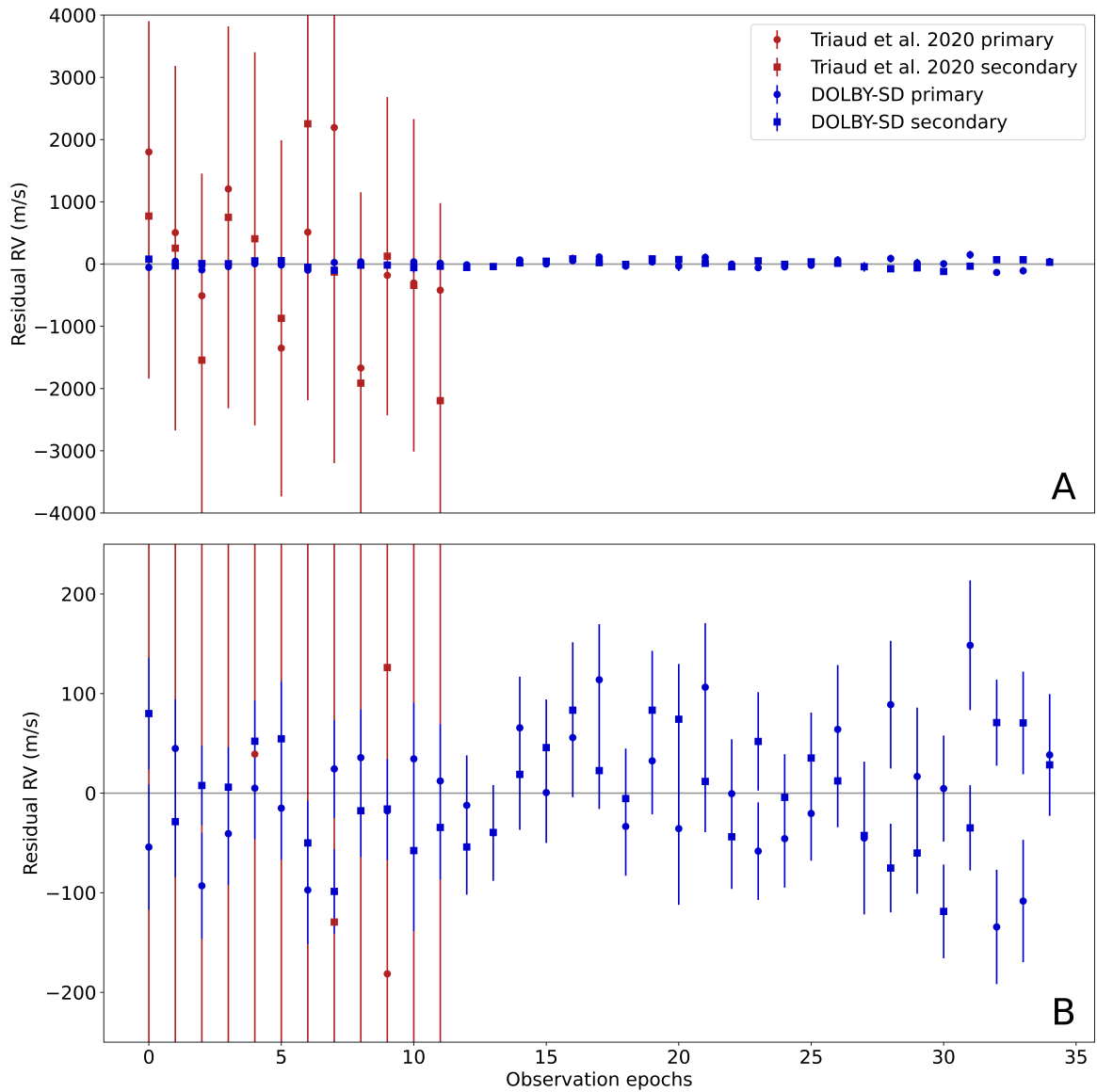}
\caption{\textbf{Comparison of radial velocity reduction methods.} Figure showing residuals relative to the first epoch in m/s for the two stars, primary (circles) and secondary (squares). The blue points represent the RVs measured using the {\tt DOLBY}-SD method (this work) and the red circles represent literature values from \cite{triaud_eclipsing_2020}. Panel B (bottom) is a zoom-in of panel A (top) so that the uncertainties on the {\tt DOLBY}-SD data can be seen.}\label{fig:rvcomp}
\end{figure}

We reanalyse all old and newly obtained UVES data. We reach a median radial velocity precision of \( 47~{\rm m\,s^{-1}}\). Thanks to {\tt DOLBY}-SD we improve the radial-velocity precision on the already published UVES data from $ \sim 1600~\rm m\,s^{-1}$. We compare our newly obtained radial velocities to values obtained on the old UVES data, reported in \cite{triaud_eclipsing_2020}. As shown in Fig~\ref{fig:rvcomp}, the performance of DOLBY-SD is evident, achieving a remarkable precision for brown dwarfs.  For comparison, \cite{blake_nirspec_2010} report a typical precision of $ 200~\rm m\,s^{-1}$ for L dwarfs. The radial velocities can be found in Table \ref{tab:rv_data1}-\ref{tab:rv_data2}.

\subsection*{Radial velocity analysis}\label{sec:rvanalysis}

We perform the initial radial velocity analysis using {\tt kima} \cite{faria_kima_2018}, a nested sampling powered analysis package allowing to fit for the number of orbiting objects in a system as a free parameter. Within {\tt kima}, we employ the BINARIESmodel , which includes the apsidal precession of the binary as a free parameter, $\dot \omega$, a time derivative of the argument of periastron \cite{baycroft_improving_2023}. Parameters are shown in Table \ref{tab:params}. We achieve a precision of $0.2\%$ on the components' masses (a factor of 30 better than the precision of 6.8\% from \cite{triaud_eclipsing_2020}). No planet or additional third body is detected in the radial velocities.
\\

We find apsidal precession rate of the binary to be significantly negative, with a value of \(\dot \omega_{\rm bin} = -343\pm126~{\rm "/yr}\). The posterior distribution of \(\dot{\omega}\) is shown in Fig \ref{fig:prisec}. Comparing the posterior density with a positive precession rate to that with a negative precession rate reveals a evidence for the negative solution with a confidence of \(99.7\%\). Alternatively framed in terms of Bayesian model comparison, this is a Bayes' Factor of 340. A typical value of 150 is used as a confidence threshold for Bayes' Factors corresponding to ``strong evidence" \cite{trotta_bayes_2008}. Hence our detection of retrograde apsidal precession is above the standard confidence threshold.
\\

We perform a N-body fit with the N-body package {\tt rebound} \cite{rein_rebound_2012} using an Integrator with Adaptive Step-size (IAS15) \cite{rein_ias15_2015} to simulate a radial velocity time-series that is compared to the observed data, and the MCMC algorithm \textit{emcee} \cite{foreman-mackey_emcee_2013} to explore parameter space. We fit a 3-body model with the third body initialised to a circular orbit, but with a range of initial inclinations. There is thus no prior weight given to polar vs coplanar orbits. The algorithm explores the parameter space, sampling from the posterior distribution, from which physical and orbital parameters are extracted.

\subsection*{Derivation of spin-induced precession equations}

In this section we derive the spin-induced precession rate in the observer's frame. This is an extension of the derivation in the appendix of \cite{baycroft_improving_2023} where spin alignment had been assumed. We relax that assumption such that the Hamiltonian from the rotational effect is \cite{goldreich_history_1966}

\begin{equation}\label{eq:ham}
    \mathcal{H} = - \frac{C_{r,0}P_2(\cos{\theta_0})+C_{r,1}P_2(\cos{\theta_1})}{(1-e^2)^{3/2}},
\end{equation}
where 

\begin{equation}
    C_{r,i} = \frac{\mathcal{G}m_0m_1J_{2,i}R_i^2}{2a^3}.
\end{equation}

Here, \(\mathcal{G}\) is the gravitational constant, \(m_i\) and \(R_i\) the mass and radius of the two components of the binary, \(a\) the total semi-major axis of the binary orbit, \(P_2\) is the second order Legendre polynomial \(P_2(x) = \dfrac{3x^2-1}{2}\), and 

\begin{equation}
    J_{2,i} = k_{2,i}\frac{\Omega_i^2R_i^3}{3\mathcal{G}m_i},
\end{equation}
where \(\Omega_i\) and \(k_{2,i}\) are the rotation rate and second Love number for body \(i\) respectively. The angle \(\theta_i\) is the obliquity, i.e. the angle between the spin axis of star \(i\)  and the normal to the orbit. It can be expressed in the observer's frame as \cite{correia_secular_2016} 

\begin{equation}
    \cos{\theta_i} = \cos{I}\cos{\varepsilon_i} + \sin{I}\sin{\varepsilon_i}\cos(\Omega - \varphi_i),
\end{equation}
where \(I\) and \(\Omega\) are the orbital inclination and longitude of ascending node of the binary orbit, \(\varepsilon_i\) is the angle between the equator of star \(i\) and the plane of the sky, and \(\varphi_i\), the spin precession angle, is the angle between the the x reference axis in the plane of the sky and the line of nodes between this plane and the equator of star \(i\).

We use the Lagrange Planetary Equations \cite{murray_solar_1999}

\begin{equation}
    \frac{d\omega}{dt} = - \frac{(1-e^2)}{eG_1}\frac{\partial\mathcal{H}}{\partial e} + \frac{\cot{I}}{G_1}\frac{\partial\mathcal{H}}{\partial I},
\end{equation}
where \(G_1\) is the norm of the orbital angular momentum 

\begin{equation}
    G_1 = \frac{m_0m_1}{m_0+m_1}\sqrt{\mathcal{G}(m_0 + m_1)a(1-e^2)}.
\end{equation}

Substituting in the Hamiltonian from eq. (\ref{eq:ham}), we get that the spin-induced apsidal precession rate is

\begin{equation}\label{eq:wdot}
\begin{split}
    \frac{d\omega}{dt} = \,&3\frac{C_{r,0}P_2(\cos{\theta_0}) + C_{r,1}P_2(\cos{\theta_1})}{G_1(1-e^2)^{3/2}} \\ &- \frac{\cot{I}}{G_1(1-e^2)^{3/2}}\left[C_{r,0}\frac{\partial P_2(\cos{\theta_0})}{\partial I} +C_{r,1}\frac{\partial P_2(\cos{\theta_1)}}{\partial I}\right],
\end{split}
\end{equation}

and 

\begin{equation}
    \frac{\partial P_2(\cos \theta_i)}{\partial I} = 3\cos\theta_i\left[\cos{I} \sin{\varepsilon_i}\cos(\Omega - \varphi_i) - \sin{I}\cos{\varepsilon_i}\right].
\end{equation}


\clearpage 

%
\bibliography{S1510} 
\bibliographystyle{sciencemag}

%
%
%
%
%
%


\section*{Acknowledgments}
We thank the anonymous referees for their helpful comments.
We acknowledge receiving observations from the European Southern Observatory (ESO), under Prog.ID 299.C-5046, 2100.C-5024, 0103.C-0042 and 111.24ZA.001. We thank the kind staff at ESO for collecting our observations.
\paragraph*{Funding:}
AHMJT acknowledges funding from the European Research Council (ERC) under the European Union's Horizon 2020 research and innovation programme (grant agreement 803193/BEBOP), by the Leverhulme Trust (research project grant RPG-2018-418), and from the ERC/UKRI Frontier Research Guarantee programme (EP/Z000327/1/CandY).
ACMC acknowledges support from the FCT, Portugal, through the CFisUC projects UIDB/04564/2020 and UIDP/04564/2020, with DOI identifiers 10.54499/UIDB/04564/2020 and 10.54499/UIDP/04564/2020, respectively.

\paragraph*{Author contributions:}
TAB developed methods and software used and perfomed the data analysis. AHMJT and LS led the observing proposals that obtained the UVES data. LS developed software, and curated the data reducing the spectra and calculating the radial velocities. TAB, LS and AHMT were involved in the inital investigation. ACMC and TAB derived the spin-induced precession equations, and studied the orbital dynamics of the system. AHMJT and ACMC acquired funding and performed validation. AHMJT supervised the work done. All authors contributed to conceptualising the experiment and to the methodology used and the vizualisation of data and results. The project administration was preformed by TAB and AHMJT. TAB wrote the majority of the manuscript with all authors contributing.
\paragraph*{Competing interests:}
There are no competing interests to declare.
\paragraph*{Data and materials availability:}
All data needed to evaluate the conclusions in the paper are present in the paper and/or the Supplementary Materials.
The radial velocity data are included in the manuscript in Tab.~\ref{tab:rv_data1}-\ref{tab:rv_data2}. The spectra are available on the ESO archive.
The codes utilised in this work are all publicly available.\\ {\tt kima}: https://www.kima.science/\\{\tt rebound}: https://rebound.readthedocs.io/en/latest/ \\{\tt emcee}: https://emcee.readthedocs.io/en/stable/




\clearpage


\begin{table}
    \centering
    \caption{\textbf{Table of parameters for the binary brown dwarf.} Parameters from both the Keplerian (\(+\)precession) fit and from the n-body fit. \(1\sigma\) uncertainties are reported in brackets precise to the last two significant digits. \(^1\)N-body fit parameters are osculating parameters at reference time bjd 2\,458\,984.600486. \(^2\) Inclination value taken from \cite{triaud_eclipsing_2020}. \(^3\) Masses for keplerian fit are derived from \(P,K,e,\) and \(i\).}
    \begin{tabular}{l|cc}
        \hline
        Parameter \& units & Keplerian fit & N-body fit\footnotemark[1] \\
        \hline
        \(P_{\rm bin}\) (days) & \(20.907495(88)\) & \(20.90769(28)\) \\
        \(e_{\rm bin}\) & \(0.36035(51)\) & \(0.35957(67)\) \\
        \(\omega_{\rm bin}\) (deg) & \(284.33(13)\) & \(284.36(11)\) \\
        \(T_{\rm peri}\) (BJD) & \(2\,459\,070.5568(73)\) & \(2\,459\,070.5601(70)\) \\
        \(\dot{\omega}_{\rm bin}\) (\({\rm "~yr^{-1}}\)) & \(-343(126)\) & --- \\
        \(i_{\rm bin}\) (deg) & \(88.5(0.1)\footnotemark[2]\) & \(88.5(0.1)\footnotemark[2]\) \\
        \(K_{\rm A}\) (\({\rm km~s^{-1}}\)) & \(16.792(13)\) & --- \\
        \(q_{\rm bin}\) & \(1.0034(10)\) & --- \\
        \(M_{\rm A}\) (\({\rm M_{\odot}}\)) & \(0.033104(68)\footnotemark[3]\) & \(0.033101(73)\) \\
        \(M_{\rm B}\) (\({\rm M_{\odot}}\)) & \(0.033219(67)\footnotemark[3]\) & \(0.033212(69)\) \\
        \hline
    \end{tabular}
    
    \label{tab:params}
\end{table}

\begin{table}
    \centering
    \caption{\textbf{Radial velocity data.} Radial velocity observations taken with UVES between August 2017 - April 2019. RV extraction performed with {\tt DOLBY}-SD.}
    \begin{tabular}{c|c|c|c|c}
    \hline
    Times &     RV$_1$ &  RVerr$_1$ &       RV$_2$ &  RVerr$_2$ \\
    \hline
    (BJD) & ${\rm m~s^{-1}}$ & ${\rm m~s^{-1}}$ & ${\rm m~s^{-1}}$ & ${\rm m~s^{-1}}$ \\
    \hline
    2457981.558 & -23.106923 & 0.056350 &  -0.613241 & 0.056057 \\2457983.531 &  -4.406202 & 0.040391 & -19.255910 & 0.056296 \\2457984.532 &   3.761294 & 0.045071 & -27.494813 & 0.039901 \\2457986.559 &   5.374921 & 0.043314 & -29.052112 & 0.040323 \\2457987.537 &   2.988056 & 0.042744 & -26.582181 & 0.040730 \\2457989.558 &  -2.997553 & 0.043627 & -20.635972 & 0.057756 \\2457991.534 &  -8.681811 & 0.046340 & -15.158845 & 0.042329 \\2458152.334 &   5.912090 & 0.040125 & -29.627186 & 0.042429 \\2458163.834 & -20.875167 & 0.039464 &  -2.845042 & 0.046541 \\2458172.846 &   4.554757 & 0.040630 & -28.234090 & 0.050263 \\2458195.825 &   5.034781 & 0.048589 & -28.702016 & 0.080853 \\2458198.845 &  -3.593771 & 0.049554 & -20.103728 & 0.052319 \\2458588.224 & -20.862209 & 0.041492 &  -2.942019 & 0.047969 \\2458588.243 & -20.753037 & 0.039004 &  -3.063875 & 0.039198 \\2458588.265 & -20.486708 & 0.042863 &  -3.165794 & 0.055679 \\2458588.284 & -20.410925 & 0.041825 &  -3.279154 & 0.048291 \\
    \hline
    \end{tabular}

    \label{tab:rv_data1}
\end{table}

\begin{table}
    \centering
    \caption{\textbf{Radial velocity data continued.} Radial velocity observations taken with UVES between May 2019 - August 2023. RV extraction performed with {\tt DOLBY}-SD.}
    \begin{tabular}{c|c|c|c|c}
    \hline
    Times &     RV$_1$ &  RVerr$_1$ &       RV$_2$ &  RVerr$_2$ \\
    \hline
    (BJD) & ${\rm m~s^{-1}}$ & ${\rm m~s^{-1}}$ & ${\rm m~s^{-1}}$ & ${\rm m~s^{-1}}$ \\
    \hline
    2458608.228 & -25.351097 & 0.052723 &   1.735621 & 0.068176 \\2458612.145 &   5.589511 & 0.048110 & -29.095287 & 0.038565 \\2458614.158 &   4.526084 & 0.040668 & -28.210611 & 0.050178 \\2458626.097 & -25.371938 & 0.045630 &   1.733123 & 0.059567 \\2458631.143 & -10.272900 & 0.070999 & -13.387907 & 0.055523 \\2458633.165 &   5.892015 & 0.057684 & -29.415012 & 0.050951 \\2460063.308 & -13.600389 & 0.046803 & -10.155635 & 0.052327 \\2460076.259 &   6.356078 & 0.040115 & -30.001234 & 0.049416 \\2460076.280 &   6.375967 & 0.040250 & -30.064740 & 0.043314 \\2460076.295 &   6.405569 & 0.038149 & -30.029529 & 0.045650 \\2460105.150 & -13.603015 & 0.058107 & -10.032466 & 0.046587 \\2460108.154 & -20.791822 & 0.071329 &  -3.033409 & 0.066947 \\2460132.984 & -27.027480 & 0.057553 &   3.280402 & 0.044459 \\2460139.089 &   6.451245 & 0.063070 & -30.133414 & 0.040886 \\2460146.995 & -13.737078 & 0.045084 & -10.089194 & 0.047018 \\2460147.010 & -13.630654 & 0.058699 &  -9.967987 & 0.042765 \\2460170.042 & -19.002846 & 0.050005 &  -4.791411 & 0.043216 \\2460171.022 & -21.174264 & 0.054583 &  -2.602333 & 0.051557 \\2460181.984 &   5.098614 & 0.054269 & -28.675622 & 0.038783\\
    \hline
    \end{tabular}

    \label{tab:rv_data2}
\end{table}

\clearpage



\renewcommand{\thefigure}{S\arabic{figure}}
\renewcommand{\thetable}{S\arabic{table}}
\renewcommand{\theequation}{S\arabic{equation}}
\renewcommand{\thepage}{S\arabic{page}}
\setcounter{figure}{0}
\setcounter{table}{0}
\setcounter{equation}{0}
\setcounter{page}{1} 














\end{document}